\documentclass[%
 reprint,
 superscriptaddress,
 amsmath,amssymb,
 aps,
]{revtex4-2}

\usepackage[utf8]{inputenc}
\usepackage[british]{babel}
\usepackage[T1]{fontenc}
\usepackage{amsmath}
\usepackage{amsfonts}
\usepackage{amssymb}
\usepackage{bm}
\usepackage{graphicx}
\usepackage{multirow}
\usepackage{tabularx}
\usepackage{booktabs}
\usepackage{siunitx}
\usepackage{makecell}
\usepackage{placeins}
\usepackage[hidelinks]{hyperref}

\newcommand{\eBP}{\hat{e}_{\mathrm{BP}}}
\newcommand{\eOSD}{\hat{e}_{\mathrm{OSD0}}}
\newcommand{\dH}{d_{\mathrm{H}}}
\newcommand{\fesc}{f_{\mathrm{esc}}}

\begin{document}

\title{Adaptive decoding of quantum LDPC codes through decoder disagreement}

\author{Maida Wang}
\affiliation{Centre for Computational Science, University College London, UK}

\author{Peter V. Coveney}
\email{p.v.coveney@ucl.ac.uk}
\affiliation{Centre for Computational Science, University College London, UK}
\affiliation{Advanced Research Computing Centre, University College London, UK}

\date{\today}

\begin{abstract}
Accurate decoding of quantum low-density parity-check (qLDPC) codes often relies on expensive post-processing search, although decoding difficulty varies strongly between syndromes. We find that the benefit of deeper post-processing search is highly concentrated in a small subset of decoding instances, and that these instances can be identified directly from the decoder itself. We introduce an adaptive decoder based on belief propagation (BP) and ordered-statistics decoding (OSD), using the disagreement between the BP hard decision and the syndrome-consistent order-zero OSD solution as an internal risk signal to determine where deeper search is required. On a $[[144,12,12]]$ bivariate bicycle code under circuit-level depolarising noise, encoding $12$ logical qubits in $144$ data qubits with distance $12$, escalating only the highest-risk $20\%$ of instances recovers $85\%$ to $92\%$ of the improvement in logical error rate available from a full one-free-variable sweep, while reducing the mean serial decoding cost by a factor of $3.6$ relative to applying the same truncated search to every instance. At matched budget, disagreement-guided routing also outperforms routing based on the BP syndrome residual, syndrome weight and random selection. The same behaviour reappears on a structurally distinct radial qLDPC code obtained from a lifted-product construction, where escalating $20\%$ of instances recovers $87\%$ to $94\%$ of the available gain. A $Z$-memory experiment on the Quantinuum H2 trapped-ion processor, implementing both $X$-type and $Z$-type checks, further shows that the signal remains predictive under device noise. These results show that decoder-internal disagreement can expose where additional decoding effort is valuable, allowing classical computation to be concentrated on the instances most likely to benefit from it.
\end{abstract}

\maketitle

\section{\label{sec:intro}Introduction}

Fault-tolerant quantum computation is a joint quantum and classical problem. Each round of syndrome extraction produces classical data that has to be decoded, and the decoder must keep pace with the syndrome stream; when the mean decoding time exceeds the round period, the backlog grows without bound and the logical clock stalls~\cite{terhal2015review,battistel2023realtime}. This is a practical constraint at the moment. Real-time decoders now run at the microsecond scale in dedicated hardware~\cite{barber2025collision}, and decoding latency and throughput are reported alongside logical performance in current memory experiments~\cite{acharya2025belowthreshold}.

Quantum low-density parity-check (qLDPC) codes sharpen this problem. Families such as the bivariate bicycle codes reach useful distances at a small fraction of the qubit overhead of the surface code~\cite{bravyi2024bbcodes,panteleev2021degenerate,kovalev2013hyperbicycle}, but degeneracy and short cycles limit what belief propagation (BP) achieves on its own. The accuracy here comes from the ordered-statistics decoding (OSD) that runs after BP: the BP reliabilities pick an information set, a first correction is built from it, and further candidate corrections are then searched around that solution~\cite{panteleev2021degenerate,roffe2020decoding}. The order-zero stage on its own, OSD-0, returns a correction that matches the syndrome but searches no candidates. The search thus can dominate the cost of accurate BP-OSD decoding~\cite{wolanski2024ambiguity}.

In the previous methods, the amount that needs to be searched is normally decided before they start. Every syndrome that reaches the post-processing stage is given the same number of candidates to try, even though some shots are far harder to decode than others. Most work on decoding cost lowers it with a recipe applied to every shot: improved BP that removes the need for post-processing altogether~\cite{muller2025relaybp}; local predecoders and hierarchical schemes that resolve the easy majority before a global decoder is called~\cite{delfosse2020hierarchical,smith2023predecoder,chamberland2023combining,knapen2026arqade}; and window-based methods that parallelise decoding in time~\cite{skoric2023parallelwindow,tan2023slidingwindow}. All of these make the pipeline cheaper as a whole, while each individual shot still receives the same treatment.

Several per-shot decisions have been developed recently, although the ones proposed so far are of a different kind. Standard BP-OSD makes one already: post-processing runs only when BP fails to converge~\cite{roffe2020decoding}. Consisting of a yes-or-no decision, it settles whether post-processing happens at all. Layered schemes send hard shots to a more capable decoder, either by measuring how far an ensemble of decoders agrees and re-running the flagged shots with a larger ensemble~\cite{shutty2024ensembling}, or by reading a confidence score off a weak decoder and switching to a stronger decoder of a different kind~\cite{toshio2025switching}. A related line uses a per-shot reliability estimate to throw shots away rather than decode them again, by aborting or post-selecting~\cite{meister2024soft,smith2024mitigating,lee2026postselection}. Reliability estimation for qLDPC codes is developing quickly along several further axes, including post-selection under reweighted error models~\cite{xie2026postselection}, prediction of BP convergence from syndrome structure~\cite{pakhunov2026convergence}, learned uncertainty estimates from neural decoders~\cite{mi2025uncertainty}, and probabilistic certificates of decoder optimality~\cite{krishnamoorthy2026certified}. In each case, the decision concerns whether a shot is accepted, whether BP converged, whether the answer can be certified, or which decoder to apply. The depth of the post-processing search itself is generally left as a fixed setting of the decoder, and how it might instead be chosen shot by shot has so far received little attention.

We are motivated by a different question here: once the fast decoder has produced an answer, how much extra search should this particular shot receive? We discovered that what is needed to answer is already in the fast decoder's own output. BP returns a bit-by-bit guess $\eBP$, taken from the signs of its posterior log-likelihood ratios. OSD-0 returns a correction $\eOSD$ built to match the observed syndrome exactly, $H\eOSD = s$, working from the reliability ordering that BP itself supplies. The two put the same syndrome to different uses, one probabilistic and one algebraic, and we take the number of positions where they disagree,
\begin{equation}
  \dH(\eBP,\eOSD) = \left\lVert \eBP \oplus \eOSD \right\rVert_0
  \label{eq:dh}
\end{equation}
as a measure of how risky that shot is. Both vectors are already in hand, so $\dH$ costs one exclusive-or and one bit count. In the experiments reported below, shots on which BP's preferred answer has to be rewritten heavily before it matches the syndrome are much more likely to be ones the fast decoder gets wrong, and they are where further search pays off.

BP and OSD-0 disagree because what BP finds most likely and what the detectors will admit do not always point to the same correction. Counting the syndrome bits tells us how much error activity a shot produced; $\dH$ tells us how much of BP's answer had to change before it fitted the code. We therefore read $\dH$ as an internal measure of the mismatch between probabilistic inference and the algebra of the code, or a constraint-induced reconstruction distance. We use it to set how much post-processing search each syndrome receives, $K = K(s)$: low-risk shots stop at OSD-0, and high-risk shots go on to a one-free-variable (1FV) search. Decoding effort becomes something handed out shot by shot rather than a fixed property of the pipeline.

We test this on two bivariate bicycle codes under circuit-level depolarising noise, on a radial qLDPC code obtained from a lifted-product construction~\cite{panteleev2022liftedproduct,scruby2026radial}, and on a memory experiment run on the Quantinuum H2 trapped-ion processor~\cite{moses2023h2}. On the $[[144,12,12]]$ code, the disagreement separates the shots the fast path decodes correctly from the ones it fails on, and sending only the highest-risk $20\%$ of shots to the stronger search recovers most of the improvement a full sweep would give, at a small fraction of its serial cost. At the same budget it beats routing on the residual left by the BP hard decision, on the syndrome weight, and at random. The $[[72,12,6]]$ code separates the two things the size of that gain depends on: how well the signal ranks risk, and how much the stronger search can fix that the fast path cannot. Both the risk ranking and the advantage under matched-budget routing reappear on the radial code, which places the mechanism outside the bivariate bicycle family. On Quantinuum hardware, the signal remains predictive under the noise of the device itself.

The central finding here is that the benefit of deeper post-processing search is concentrated in a small subset of decoding instances, and that this variation can be read directly from the fast decoder itself. The BP-OSD disagreement provides the corresponding risk signal: it is produced inside a single BP-OSD pass, with no ensemble, no second decoder and no trained confidence model, and it sets the depth of the search that follows within the same decoding hierarchy. Adaptive decoding and confidence-guided routing themselves have prior art, cited above; to our knowledge, BP-OSD decision disagreement has not previously been used to control instance-wise OSD search effort in qLDPC decoding.

Section~\ref{sec:method} defines the disagreement signal and the adaptive policy. Section~\ref{sec:risk} establishes that disagreement tracks decoding risk, Sec.~\ref{sec:escalation} shows what risk-guided escalation buys, Sec.~\ref{sec:pareto} quantifies the trade-off between accuracy and computation, the role of decoder headroom and the transfer to another code family, and Sec.~\ref{sec:hardware} reports the hardware validation. Section~\ref{sec:discussion} discusses implications and limitations, and Sec.~\ref{sec:methods} collects methodological detail.

\section{\label{sec:method}Adaptive decoding by decoder disagreement}

\begin{figure*}[!tbp]
\centering
\includegraphics[width=0.95\textwidth]{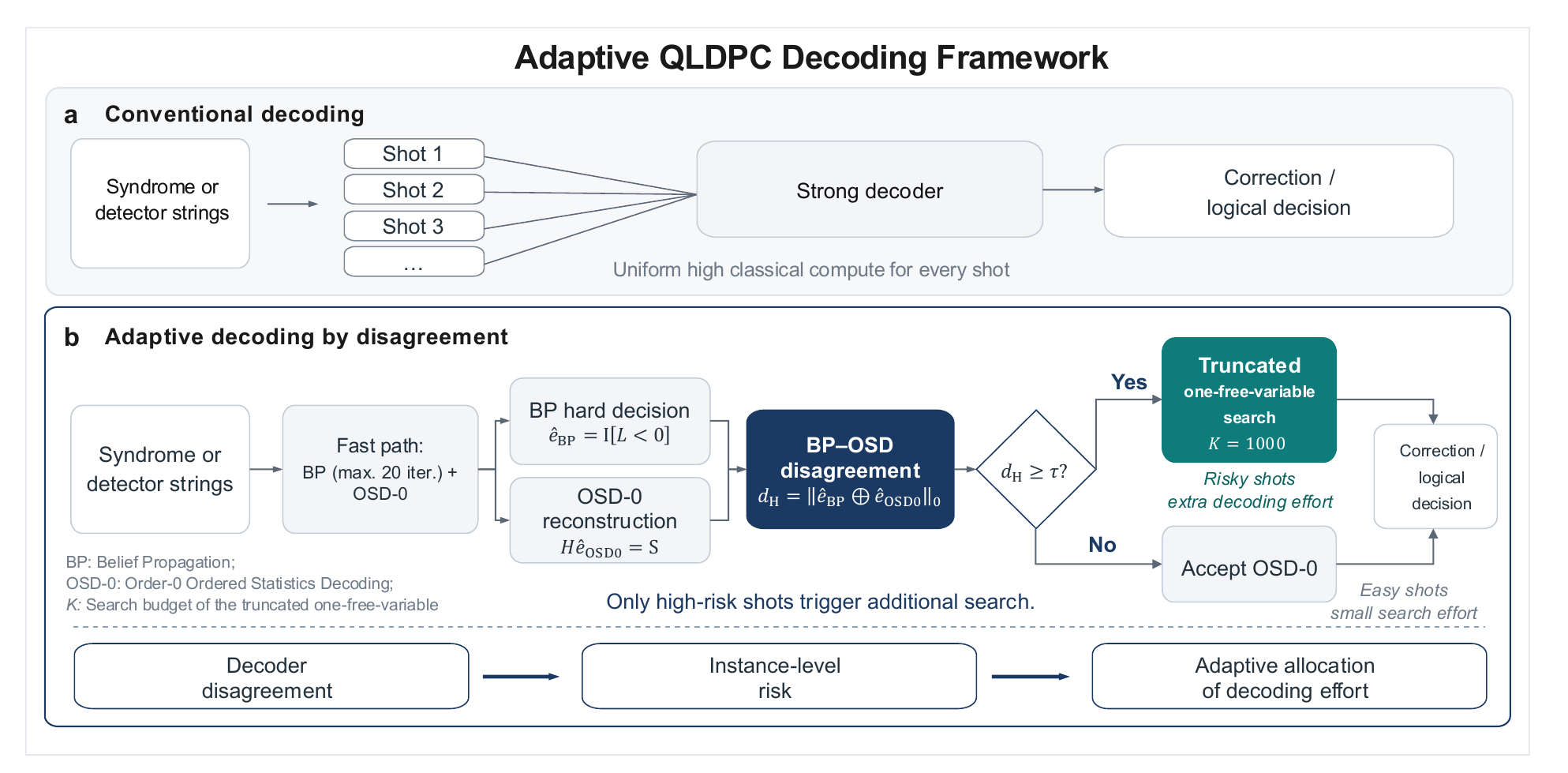}
\caption{\label{fig:framework}Adaptive qLDPC decoding by BP-OSD disagreement. (a) Conventional decoding applies the same procedure and the same search budget to every shot, so the classical cost of accurate decoding scales with the entire syndrome stream. (b) In the adaptive pipeline every shot first takes the fast path, BP with at most $20$ iterations followed by OSD-0. That pass yields two estimates of the same error: the probabilistic BP hard decision $\eBP = \mathbb{I}[L<0]$ and the syndrome-consistent output $\eOSD$ satisfying $H\eOSD = s$, which coincides with $\eBP$ whenever BP already satisfies the syndrome. Their Hamming disagreement $\dH = \lVert \eBP \oplus \eOSD \rVert_0$ measures how risky the shot is, at negligible extra cost. Shots with $\dH \geq \tau$ are escalated to a truncated 1FV sweep with budget $K = 1000$; the remainder accept the OSD-0 output. Additional decoding effort is allocated where the decoder itself flags elevated risk.}
\end{figure*}

\subsection{\label{sec:two-estimates}Two estimates inside one fast decoder}

Decoding is performed on a detector error model derived from the circuit-level noise model~\cite{gidney2021stim}, written as a detector check matrix $H$ over error-mechanism coordinates with prior probabilities $p_i$, a vector of detector events $s$, which we refer to throughout as the syndrome, and a matrix of logical observables. BP is run in the normalised min-sum formulation~\cite{roffe2020decoding,roffe2022ldpc} for at most $20$ iterations and returns posterior log-likelihood ratios $L_i$, from which the hard decision
\begin{equation}
  \eBP = \mathbb{I}\!\left[L < 0\right]
\end{equation}
is read bitwise. This is the error pattern BP considers most likely, taken one mechanism at a time. It has been demonstrated that there is no compulsion for the syndrome to be reproduced, and in instances where the BP has not converged, this is typically the case.

OSD-0 uses the same posteriors differently. The BP-derived reliability ordering picks an information set of independent columns, and a correction is then built algebraically over $\mathbb{F}_2$ so that it matches the observed syndrome exactly,
\begin{equation}
  H\,\eOSD = s ,
\end{equation}
by construction. BP answers which combination of error mechanisms is most likely, mechanism by mechanism; OSD-0 answers which combination is consistent with every detector constraint under that ranking.

The fast path returns a syndrome-consistent output, which we write as $\eOSD$ throughout. When the BP hard decision already satisfies the syndrome, that output is the BP solution itself. Otherwise, OSD-0 uses the BP-derived reliability ordering and an algebraic reconstruction to produce a correction that does. The disagreement of Eq.~\eqref{eq:dh} is therefore zero exactly when BP already supplies the syndrome-consistent answer, and otherwise counts how much had to be rebuilt to match the syndrome.

\subsection{\label{sec:signal}Disagreement as a risk signal}

Both vectors come out of the decoding path anyway, so $\dH$ costs one exclusive-or and one bit count. The comparison is useful because the two estimates are built under different rules. The vector $\eBP$ is assembled from the posteriors one mechanism at a time. The syndrome has already shaped those posteriors through the BP iterations, but nothing in the bitwise decision requires the result to reproduce it. The fast-path output has to match it, and pays for that by taking whatever the reliability ordering gives outside the information set. The gap between them is the amount of rewriting that matching the syndrome costs, which is why we call $\dH$ a constraint-induced reconstruction distance. A large $\dH$ therefore says that BP's posteriors and the checks of the code disagree over many error mechanisms at once. It is not a physical observable, and it is not the distance to the nearest codeword: OSD-0 builds a correction guided by reliabilities, it does not find the closest one.

Two simpler quantities come out of the same pass, and we use them as comparators throughout. The syndrome weight $w_s = \lVert s \rVert_0$ counts how many detectors fired. The BP residual
\begin{equation}
  r_{\mathrm{BP}} = \left\lVert H\eBP \oplus s \right\rVert_0
  \label{eq:res}
\end{equation}
counts the detector constraints that the BP hard decision still violates. The three ask progressively sharper questions: how much error activity was seen, how many checks BP still fails, and how much of BP's answer has to be rebuilt before it stops failing them. A shot can set off many detectors that BP explains without difficulty, and a shot with few detectors can still force the reconstruction to overrule BP in many places.

\subsection{\label{sec:policy}Instance-adaptive search depth}

The adaptive decoder is fixed in advance and works as follows. Every shot takes the fast path: BP with at most $20$ iterations followed by OSD-0, and $\dH$ is computed. If $\dH$ falls below the routing cutoff, we keep the fast-path answer. If it is at or above the cutoff, the shot goes to a truncated 1FV sweep, which flips the first $K$ free variables of the BP-derived combination-sweep ordering one at a time and scores each resulting candidate by the channel weight
\begin{equation}
  S(x) = \sum_{i\,:\,x_i = 1} \log \frac{1}{p_i} ,
\end{equation}
evaluated on the detector error model channel probabilities, and the lowest-scoring candidate is returned. The score adds up the prior weight of every mechanism a candidate flips. It stands in for the candidate likelihood and is not the exact independent-Bernoulli negative log-likelihood. The primary branch uses $K = 1000$, with $K = 500$ examined in Appendix~\ref{app:extended}.

As the strong-search reference, we take the full 1FV sweep, which evaluates every single-free-variable perturbation in the OSD combination-sweep candidate family~\cite{roffe2020decoding}. Because that sweep can be stopped after any number of candidates, its depth is a natural quantity to adapt, and to our knowledge $\dH$ has not been used before to set it shot by shot. The full 1FV sweep is a controlled strong-search reference and should not be read as an exhaustive search over physical error configurations.

Only the search budget changes. Writing $k_{\mathrm{free}}$ for the number of free columns, which is $10446$ for the $[[144,12,12]]$ detector error model used here, the strong-search reference corresponds to a constant budget $K = k_{\mathrm{free}}$, while the adaptive policy sets
\begin{equation}
  K = K(s) \in \{0,\,1000\}
  \label{eq:budget}
\end{equation}
shot by shot. BP itself, the choice of information set, the scoring function and the decoding objective all stay as they are. Instead of making the search longer for everyone, we give each shot a search length that follows how far its two estimates disagree.

\subsection{\label{sec:cost}Cost model}

For an escalated fraction $\fesc$ of instances, the mean serial decoding cost per shot is
\begin{equation}
  \bar{T} = \mathbb{E}\!\left[T_{\mathrm{fast}}\right] + \fesc\,\mathbb{E}\!\left[\Delta T_K \mid \text{escalated}\right],
  \label{eq:cost}
\end{equation}
with $T_{\mathrm{fast}}$ the cost of the fast path and $\Delta T_K$ the additional cost of the escalated branch. The conditioning matters. The shots the router picks are not average shots, and their search costs a little more than the mean, so simply mixing the two fixed costs slightly underestimates what the adaptive policy actually costs. In a running system the expensive branch then has to keep up with a fraction $\fesc$ of the incoming syndromes rather than all of them.

\section{\label{sec:risk}Decoder disagreement shows which shots are at risk}

\begin{figure*}[!tbp]
\centering
\includegraphics[width=0.95\textwidth]{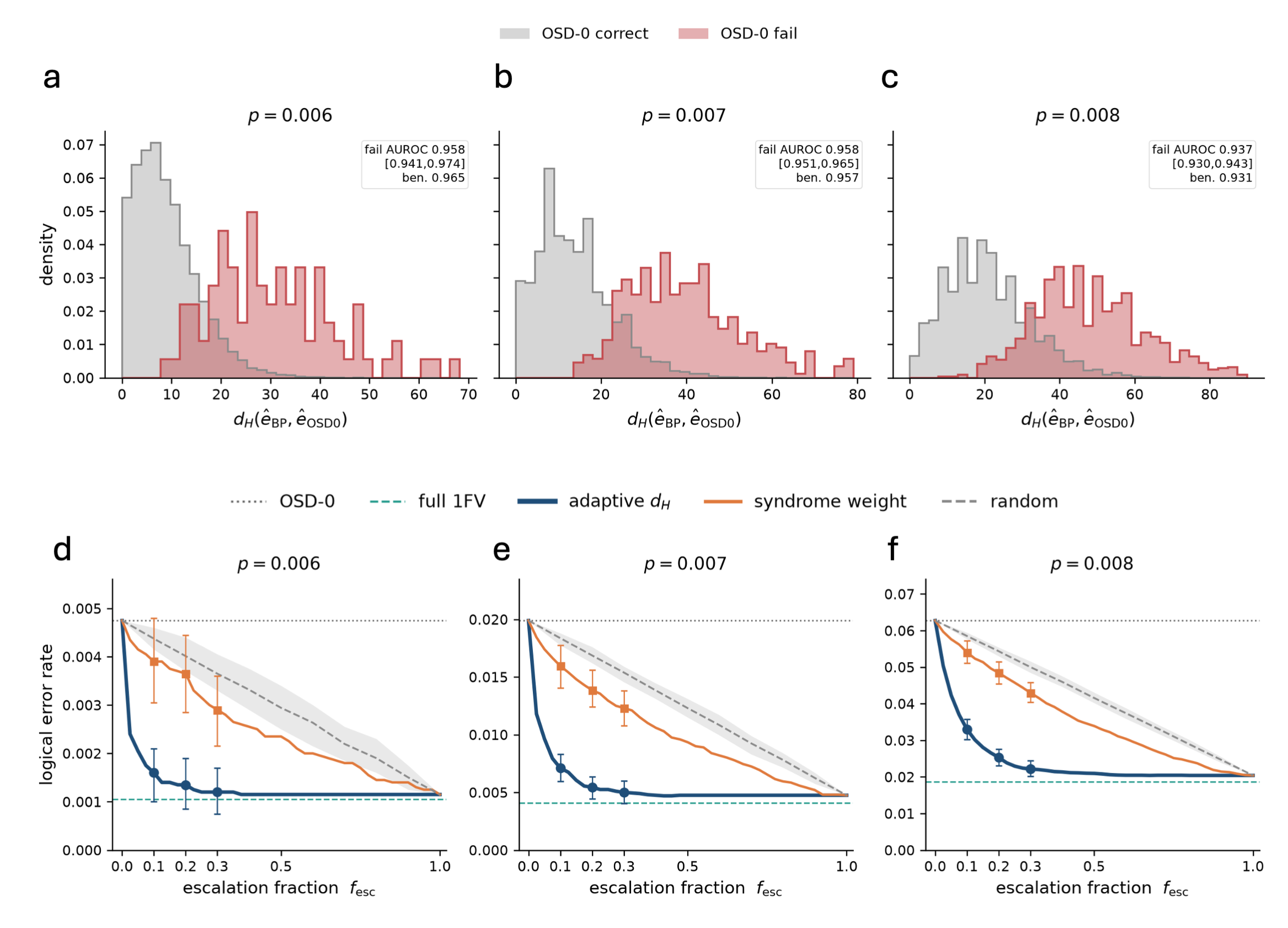}
\caption{\label{fig:risk}Disagreement tracks decoding risk and enables a targeted improvement in logical error rate, for the $[[144,12,12]]$ bivariate bicycle code over $24$ syndrome extraction rounds with $20{,}000$ shots per operating point. (a to c) Distributions of the BP-OSD disagreement $\dH$ for shots on which OSD-0 succeeds and fails, at $p = 0.006$, $0.007$ and $0.008$. Insets report the area under the receiver operating characteristic curve (AUROC) for OSD-0 failure with its bootstrap $95\%$ interval, together with the corresponding value for beneficial escalation (labelled \emph{ben.} in the panels), the case in which OSD-0 fails while the full 1FV sweep succeeds. (d to f) Logical error rate against escalation fraction $\fesc$ for routing on disagreement, on syndrome weight, and at random, shown against the OSD-0 baseline (dotted) and the full 1FV reference (dashed); the random curve interpolates between the baseline and the always-on $K = 1000$ branch. The BP residual comparator is tabulated in Appendix~\ref{app:controls}. Error bars and the shaded random-routing band are bootstrap $95\%$ intervals; random routing is averaged over $200$ seeds.}
\end{figure*}

The first thing to check is whether the disagreement between the two fast-path estimates says anything about how decoding turns out. The primary benchmark is the $[[144,12,12]]$ bivariate bicycle code~\cite{bravyi2024bbcodes} over $24$ rounds of syndrome extraction under uniform circuit-level depolarising noise, sampled at $p = 0.006$, $0.007$ and $0.008$ with $20{,}000$ shots per operating point. A shot counts as a logical failure if any of the twelve logical $Z$ observables is misidentified.

Figure~\ref{fig:risk}(a to c) shows the distribution of $\dH$ separately for shots on which OSD-0 succeeds and shots on which it fails. The two are well separated at all three noise rates. Measured by the AUROC for OSD-0 failure, the separation is $0.958$ $[0.941, 0.974]$, $0.958$ $[0.951, 0.965]$ and $0.937$ $[0.930, 0.943]$ at $p = 0.006$, $0.007$ and $0.008$.

In practice a slightly different set matters: the shots where OSD-0 fails and the full 1FV sweep succeeds, since those are the ones worth escalating. The $\dH$ ranking is heavily loaded with such shots, with AUROC values of $0.965$, $0.957$ and $0.931$. This says that the ranking as a whole puts them near the top. It does not say that, among the shots the fast decoder has already got wrong, a larger $\dH$ makes a rescue more likely; Sec.~\ref{sec:headroom} returns to that. With an AUROC below one, $\dH$ ranks shots by risk rather than deciding any individual case.

Standard BP-OSD already makes a yes-or-no judgement of this kind, running post-processing only when BP fails to converge~\cite{roffe2020decoding}, and that flag and $\dH$ behave very differently on this benchmark. On the $[[144,12,12]]$ benchmark that flag says almost nothing about whether OSD-0 will fail. BP converges on $9.18\%$, $3.06\%$ and $0.68\%$ of shots at the three noise rates, no OSD-0 failure occurs among converged shots, and the flag reaches AUROC values of only $0.546$, $0.516$ and $0.504$. Routing on it would send between $90.8\%$ and $99.3\%$ of shots to the expensive branch, which is not a $10\%$ to $30\%$ budget but nearly everything. Nor is $\dH$ merely another way of saying that BP failed to converge: among the non-converged shots alone it still reaches $0.954$, $0.957$ and $0.936$.

The $[[72,12,6]]$ code, decoded the same way with $10{,}000$ shots per point and scored against the same target, makes the comparison in a regime where the convergence flag is not almost always on. BP converges on $54.33\%$ of shots at $p=0.006$ and $29.31\%$ at $p=0.008$, and OSD-0 failures do occur among converged shots, $46$ and $50$ respectively. The flag is genuinely predictive there, with AUROC values of $0.724$ $[0.707, 0.739]$ and $0.648$ $[0.640, 0.655]$, but it remains well behind the disagreement signal at $0.831$ $[0.810, 0.852]$ and $0.839$ $[0.829, 0.850]$, and it would escalate $45.7\%$ and $70.7\%$ of the stream. Accuracy is not the only difference. A yes-or-no flag splits the stream at whatever rate the code and the noise happen to produce, whereas a graded score lets the budget be chosen.

A third comparator sits between these two. The residual $r_{\mathrm{BP}}$ of Eq.~\eqref{eq:res} is much closer in spirit to $\dH$ than the syndrome weight is, and the two track each other closely, with Spearman coefficients of $0.956$, $0.943$ and $0.927$ at $p = 0.006$, $0.007$ and $0.008$. They are not interchangeable. For OSD-0 failure the residual reaches AUROC values of $0.842$, $0.795$ and $0.769$, against $0.958$, $0.958$ and $0.937$ for $\dH$ and $0.654$, $0.657$ and $0.659$ for the syndrome weight. Counting the checks BP still fails is a much better guide than counting detectors that fired, and still clearly worse than measuring how far BP's answer has to move to stop failing them.

\section{\label{sec:escalation}Risk-guided search recovers most of the available gain}

Since $\dH$ picks out the difficult shots, the extra search can now be aimed at them. We rank shots by $\dH$, escalate the fraction $\fesc$ with the largest $\dH$ to the truncated branch with $K = 1000$, and accept the fast-path output on the remainder, so that the per-instance budget takes the two values of Eq.~\eqref{eq:budget}.

Figure~\ref{fig:risk}(d to f) shows the resulting logical error rate against $\fesc$. Without escalation the OSD-0 baseline gives logical error rates of $0.00475$, $0.01990$ and $0.06275$ at $p = 0.006$, $0.007$ and $0.008$, while the full 1FV reference gives $0.00105$, $0.00410$ and $0.01865$; the strong search is worth a factor of $4.52$, $4.85$ and $3.36$ in logical error rate. Escalating the $20\%$ of shots with the largest disagreement gives $0.00135$, $0.00545$ and $0.02530$, recovering $91.9\%$, $91.5\%$ and $84.9\%$ of that available improvement. At $30\%$ the recovered fractions rise to $95.9\%$, $94.3\%$ and $92.1\%$. The curves drop steeply at small $\fesc$ and then flatten, which is what happens when nearly all of the benefit of extra search sits in a small group of shots.

The improvement is not a consequence of running the stronger search on some of the shots. At the same budget, which shots are chosen is what decides the outcome. At $p = 0.007$ and $\fesc = 0.20$, routing on disagreement gives $0.00545$, on the BP residual $0.01140$, on syndrome weight $0.01385$, and at random $0.01687$. The random curve interpolates between the OSD-0 baseline and the always-on $K = 1000$ branch rather than the full sweep, since escalating a random fraction simply mixes the two, $0.8 \times 0.01990 + 0.2 \times 0.00475 = 0.01687$. In recovered terms the residual gets back $53.8\%$ of the available improvement at this budget against $91.5\%$ for $\dH$, and across all nine tested combinations of noise rate and budget $\dH$ gives the lower logical error rate. Both comparators are informative and both beat chance; neither matches the ranking $\dH$ gives. This is the practical form of the distinction drawn in Sec.~\ref{sec:signal}.

Escalation budgets in Fig.~\ref{fig:risk}(d to f) are imposed exactly, by ranking $\dH$ on the evaluation sample and taking the top $k = \mathrm{round}(\fesc n)$ shots, so that accuracy and compute are reported on identical shots. In use, the cutoff would instead be fixed on a separate stream of calibration shots. Calibrating $\tau$ on twenty independent seeds and applying it to twenty held-out seeds, with the folds then exchanged, gives $\tau = 14$, $21$ and $32$; the realised escalation fractions are $20.9\%$, $21.7\%$ and $20.8\%$, within $1.7$ percentage points of the nominal $20\%$ target, and the resulting logical error rates of $0.00135$, $0.00525$ and $0.02475$ are consistent with the in-sample routing curves at comparable budgets. A threshold calibrated on one sample therefore transfers to another; details are given in Appendix~\ref{app:controls}.
The gain also survives with a shallower $K = 500$ branch, though it recovers less at the higher noise rates because that branch can fix less; the comparison is tabulated in Appendix~\ref{app:extended}.
How much a shot stands to gain from extra search therefore varies sharply from one shot to the next, and the decoder can see that variation from the inside.

\section{\label{sec:pareto}Routing the search improves the trade-off between accuracy and computation}

\begin{figure*}[!tbp]
\centering
\includegraphics[width=0.95\textwidth]{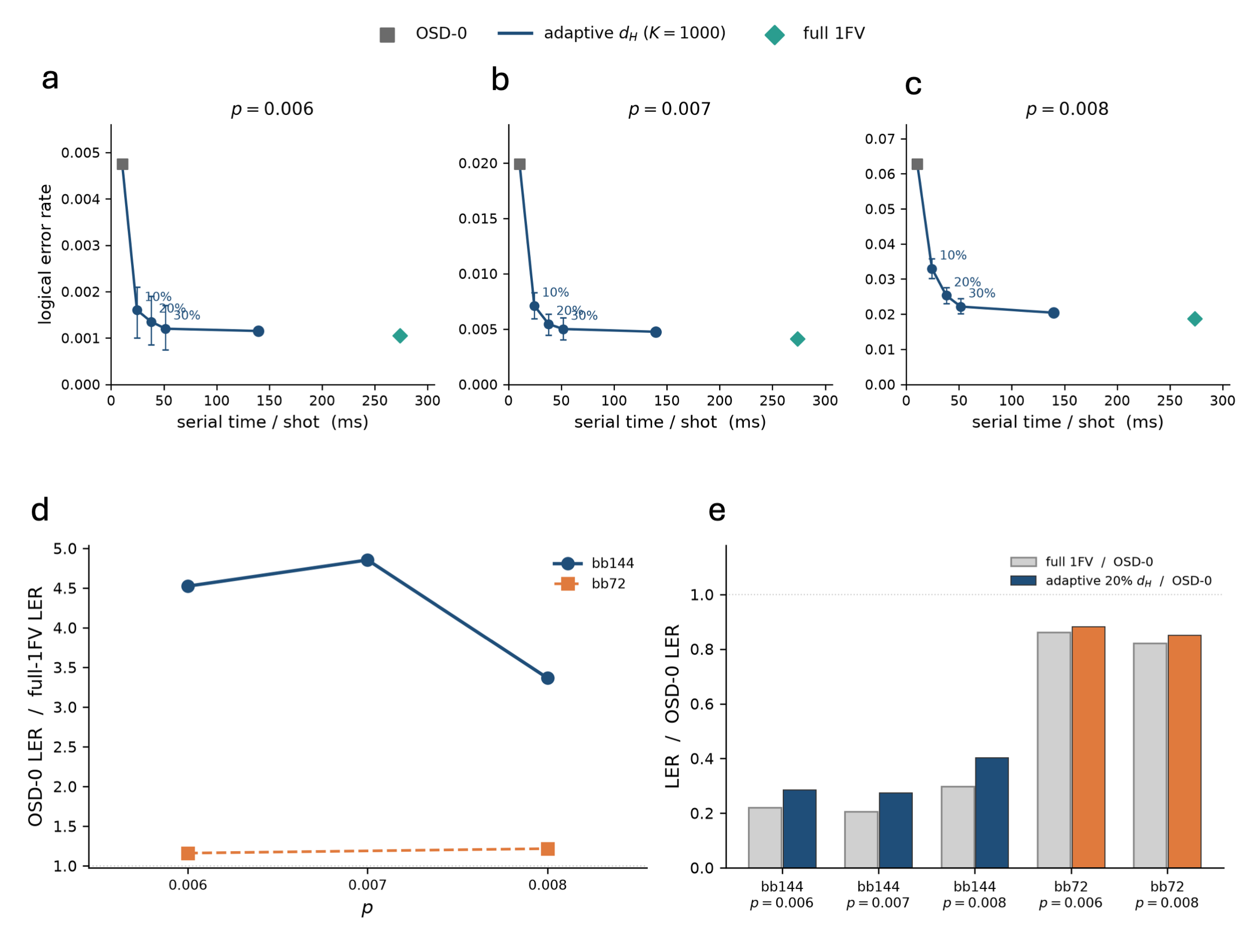}
\caption{\label{fig:pareto}Trade-off between accuracy and computation, and the role of decoder headroom. (a to c) Logical error rate against measured single-threaded serial decoding time per shot at $p = 0.006$, $0.007$ and $0.008$, for the OSD-0 fast path, the adaptive policy at escalation fractions of $10\%$, $20\%$ and $30\%$, the truncated branch applied to every shot, and the full 1FV reference. The adaptive points sit on a different part of the trade-off from the fixed strong branches, keeping most of the available improvement at a fraction of the serial cost. (d) Ratio of the OSD-0 logical error rate to the full 1FV logical error rate, the decoder headroom, for the $[[144,12,12]]$ and $[[72,12,6]]$ codes. (e) Logical error rate relative to the OSD-0 baseline for the full 1FV sweep and for escalation of the top $20\%$ by disagreement. The signal ranks risk well on both codes; what differs is how much the stronger search can fix.}
\end{figure*}

\subsection{\label{sec:timing}Measured serial decoding cost}

A smaller escalation budget is useful only if it translates into a smaller cost. Decoding time was measured on a single node in one serial process with a single OpenMP thread, with the escalated branch terminating after its $K$ candidates rather than completing the sweep, so that the truncated costs are measured rather than prorated. The per-shot time covers the complete fast path of BP with at most $20$ iterations followed by the OSD-0 stage, together with the additional search on escalated shots.

The fast path costs $10.15$~ms per shot. Applying the truncated branch to every shot costs $126.2$~ms at $K = 500$ and $131.9$~ms at $K = 1000$, and the full 1FV sweep costs $273.7$~ms. Under disagreement-guided routing the mean cost is $23.2$, $36.3$ and $49.1$~ms per shot at escalation fractions of $10\%$, $20\%$ and $30\%$, the bootstrap interval at $20\%$ running from $32.9$ to $39.9$~ms. Simply mixing the two fixed costs would predict $34.5$~ms at that budget, slightly below what we measure, because the shots the router picks run a somewhat more expensive search than average.

Two comparisons are worth keeping apart, because they measure different savings. Against the full sweep the adaptive policy at $\fesc = 0.20$ costs a factor of $7.5$ less, but part of that comes from cutting the sweep off at $K = 1000$ rather than from routing. Against the same $K = 1000$ branch run on every shot the saving is a factor of $3.6$, which is the part of the saving that routing contributes on its own. Figure~\ref{fig:pareto}(a to c) shows the resulting operating points. The adaptive policy sits on a different part of the trade-off from the fixed strong branches, keeping most of the available accuracy improvement at a small fraction of the serial cost.

The benchmark also locates the cost of escalation. At $K = 1000$ the fast path accounts for about $10.2$~ms, the algebraic preprocessing that sets the sweep up, meaning the elimination and solve described in Sec.~\ref{sec:methods}, for about $112.1$~ms, and scoring the thousand candidates for about $9.6$~ms; at $K = 500$ the scoring prefix costs only about $3.9$~ms, which is why the two truncated branches differ so little in total cost. What makes escalation expensive is entering the algebraic branch at all, not the length of the candidate prefix, so selective routing saves computation chiefly by keeping low-risk shots out of that branch.

Both figures are serial classical decoding costs measured on one node under fixed conditions, and they reflect this implementation of the fast path and the search branch.

\subsection{\label{sec:headroom}The gain needs both a good signal and something for the stronger search to fix}

How large the gain is depends on the code, and the $[[72,12,6]]$ code, decoded over $12$ rounds with $10{,}000$ shots per operating point, shows what sets it.

There the fast path gives logical error rates of $0.0406$ and $0.1339$ at $p = 0.006$ and $0.008$, and the full 1FV sweep gives $0.0350$ and $0.1101$. The strong search is worth a factor of $1.16$ and $1.22$, against $3.36$ to $4.85$ on the $[[144,12,12]]$ code, as Fig.~\ref{fig:pareto}(d) shows. The router itself still works: $\dH$ separates OSD-0 successes from failures with AUROC values of $0.831$ and $0.839$, and $20\%$ escalation gives $0.0358$ and $0.1140$, so that Fig.~\ref{fig:pareto}(e) places the adaptive policy close to the full sweep in relative terms. The routing is no less efficient on this code; there is simply much less accuracy left for it to win back.

For adaptive routing to pay off on a given code, at a given noise strength, two conditions have to hold together, and it is easy to mistake one for the other. The first is that the signal can tell which shots are at risk of a logical failure. The second is that the escalated search can actually repair those shots, which is what we mean by headroom: the gap in logical error rate between the fast path and the full sweep. A signal that ranks risk well is not on its own enough to produce a large gain. On $[[144,12,12]]$ the signal separates failures from successes well and that gap is wide, which is why the gain is large; on $[[72,12,6]]$ the signal separates them almost as well but the gap is narrow, so little accuracy is available to win back. In short, the signal decides which shots the extra computation should go to, and the strength of the escalated search decides how much logical error rate that computation buys back. The $K = 500$ comparison in Appendix~\ref{app:extended} makes the same point from the other side: holding the router fixed and weakening the escalated branch also cuts the fraction of the available improvement that is recovered, and the shortfall grows with the noise strength. This is an empirical rule, drawn from the two codes and the two branch depths studied here.

\subsection{\label{sec:radial}Cross-family transfer to a radial qLDPC code}

\begin{figure*}[!tbp]
\centering
\includegraphics[width=0.8\textwidth]{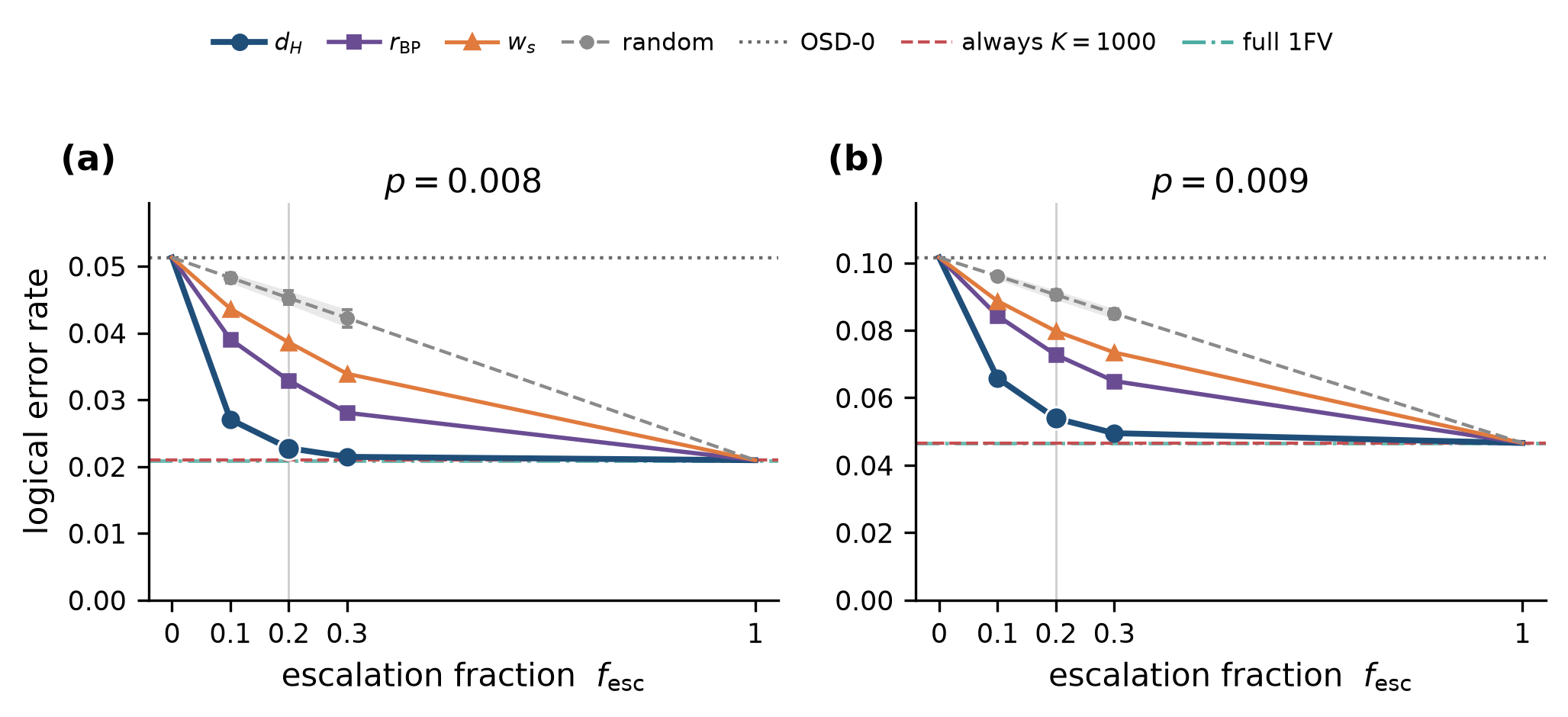}
\caption{\label{fig:radial}Transfer of disagreement-guided routing to a radial qLDPC code. Logical error rate against escalation fraction for the $n = 90$, $k = 8$ radial instance over twelve rounds of syndrome extraction at (a) $p = 0.008$ and (b) $p = 0.009$, with $20{,}000$ test shots per operating point. Exact matched-budget routing on $\dH$, on the BP residual $r_{\mathrm{BP}}$ and on the syndrome weight $w_s$ is compared with random routing, averaged over $200$ seeds; every escalated shot goes to the same $K = 1000$ branch. The OSD-0 baseline, the $K = 1000$ branch applied to every shot and the full 1FV reference are shown for comparison, the last two differ by a single logical failure at each operating point. Routing on disagreement gives the lowest logical error rate at every budget, and at $20\%$ escalation it recovers $93.9\%$ and $86.6\%$ of the improvement available from the full sweep.}
\end{figure*}

Both the mechanism and its boundary have so far been established on bivariate bicycle codes. To see how far the ordering carries, we repeat the same analysis on a radial qLDPC code, a family obtained from a lifted-product construction~\cite{panteleev2022liftedproduct}: the $n = 90$, $k = 8$ instance of Scruby, Hillmann and Roffe~\cite{scruby2026radial}, taken from their published parity-check matrices, with check weight $6$ and qubit degree $3$. The circuit is a twelve-round circuit-level $Z$-memory, decoded with the same fast path, the same candidate score and the same $K = 1000$ branch as before, and for this detector error model $k_{\mathrm{free}} = 3289$, so the branch is still a genuine truncation. The two operating points, $p = 0.008$ and $p = 0.009$, were fixed from a blind pilot that looked only at strong-search headroom, before any disagreement statistic was computed; the results below come from independent test sets of $20{,}000$ shots per point.

The fast path fails on $5.130\%$ of shots at $p = 0.008$ and $10.160\%$ at $p = 0.009$, and the full 1FV sweep brings these to $2.090\%$ and $4.655\%$, so this code sits in the high-headroom regime rather than the one the $[[72,12,6]]$ control occupies. Applying $K = 1000$ to every shot reproduces the full sweep to within a single logical failure at both points. Scored against the same OSD-0 failure target, $\dH$ reaches an AUROC of $0.951$ $[0.946, 0.956]$ and $0.940$ $[0.935, 0.944]$, the BP residual $0.832$ and $0.809$, and the syndrome weight $0.707$ and $0.701$. The ordering $\dH > r_{\mathrm{BP}} > w_s$ found on the $[[144,12,12]]$ code appears again here, on a code with a different construction and a different check geometry.

Figure~\ref{fig:radial} shows what that ordering buys under exact matched-budget routing, with every method sending the same number of shots to the same branch. At $20\%$ escalation and $p = 0.008$, routing on $\dH$ gives a logical error rate of $0.02275$, against $0.03290$ on the residual, $0.03860$ on the syndrome weight and $0.04525$ at random; at $p = 0.009$ the four values are $0.05390$, $0.07275$, $0.07975$ and $0.09054$. The $20\%$ policy takes back $93.9\%$ and $86.6\%$ of the improvement available between OSD-0 and the full sweep. The paired difference between disagreement and residual routing is $-0.01015$ $[-0.01195, -0.00880]$ at $p = 0.008$ and $-0.01885$ $[-0.02060, -0.01690]$ at $p = 0.009$, or $203$ and $377$ fewer logical failures. Since the two signals are closely related, that gap is the informative one: at the same budget, $\dH$ gathers more of the shots on which the extra search actually changes the outcome. The ordering that the decoder produces internally therefore survives a substantial change in code construction.

\section{\label{sec:hardware}Decoder disagreement on hardware}

\begin{figure*}[!tbp]
\centering
\includegraphics[width=0.9\textwidth]{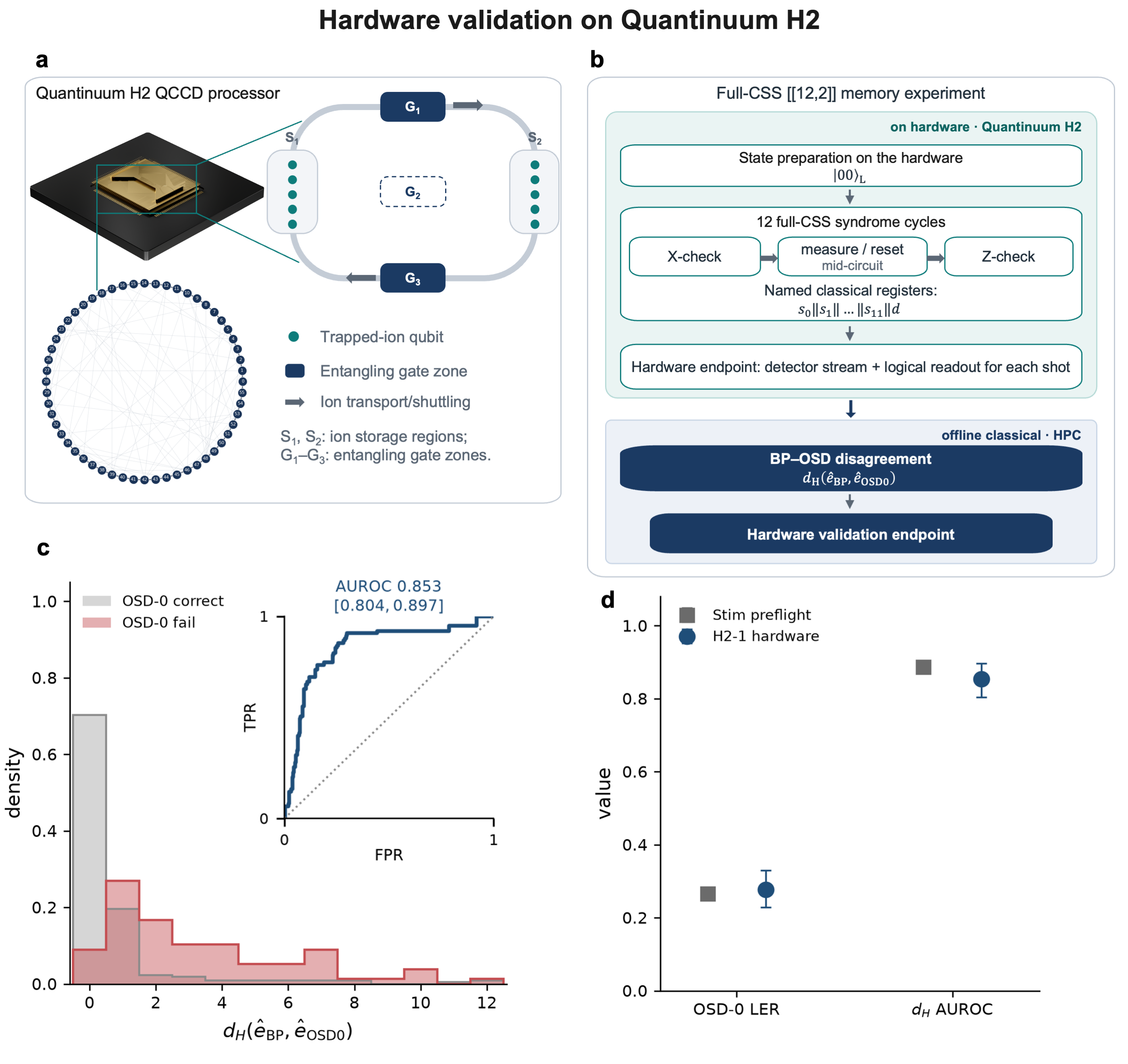}
\caption{\label{fig:hardware}Full Calderbank-Shor-Steane (CSS) memory experiment on the Quantinuum H2 trapped-ion processor. (a) The H2 quantum charge-coupled device architecture, with ion storage regions and entangling gate zones connected by ion transport, together with the connectivity graph of the experiment. (b) The $[[12,2,2]]$ full-CSS memory experiment: encoded state preparation on hardware, twelve rounds of interleaved $X$-check and $Z$-check extraction with mid-circuit measurement and reset into named classical registers, and the offline classical decoding and analysis carried out in the high-performance computing (HPC) environment, in which the detector stream is decoded and the BP-OSD disagreement computed. (c) Distributions of $\dH$ on hardware for shots on which OSD-0 succeeds and fails, with the corresponding receiver operating characteristic curve. (d) The Stim preflight and the H2 hardware run compared for the OSD-0 logical error rate and for how well $\dH$ separates failures from successes.}
\end{figure*}

Everything above was measured in simulation, where the noise follows a model we chose. Real hardware does not, so we ran a memory experiment on the Quantinuum H2 trapped-ion processor to see whether the signal still works there. The experiment uses a small instance of the bivariate bicycle construction with $(\ell,m) = (3,2)$ and $A = B = x + y$, giving parameters $[[12,2,2]]$, on $22$ physical qubits: $12$ data qubits with five $X$-check and five $Z$-check ancillas. Both CSS check types are extracted over twelve rounds, with mid-circuit measurement and reset, making this a full-CSS $Z$-memory experiment. Encoded state preparation is measurement-based, with the random first-round $X$-check outcomes absorbed into the Pauli frame rather than corrected in the circuit. No recovery operation is applied on the device; the returned measurement records were decoded and analysed offline in the HPC environment, using the same BP-OSD pipeline as the simulation study. The analysis was specified in advance, with how well $\dH$ separates OSD-0 failures from successes as the primary endpoint, and $300$ shots were collected in a single job.

Figure~\ref{fig:hardware} shows the outcome. OSD-0 fails on $83$ of $300$ shots, a logical error rate of $0.277$ $[0.229, 0.330]$, against $0.266$ in a preflight simulation of the same circuit in Stim~\cite{gidney2021stim}, so the device operated close to the intended logical error regime. The disagreement is still clearly predictive under the noise of the device itself, which is not the single-parameter circuit-level model the simulations assume: the mean $\dH$ is $1.09$ on shots decoded correctly and $4.96$ on shots that failed, and the AUROC is $0.853$ $[0.804, 0.897]$ against $0.886$ in the preflight.

The experiment validates the risk signal on hardware. Applied to every shot, the full 1FV sweep rescued three OSD-0 failures and turned three OSD-0-correct shots into logical failures, leaving the rate unchanged at $83$ of $300$; the transition counts are given in Appendix~\ref{app:controls}. The absence of a net accuracy gain is consistent with the very limited strong-search headroom of this configuration, in which the distance-two code offers essentially no guaranteed single-error correction. It also makes the point of Sec.~\ref{sec:headroom} concrete: knowing which shots are risky is not the same as knowing which ones a stronger search can fix.

\section{\label{sec:discussion}Discussion}

Decoding effort is usually set once and then applied to every shot. The results here suggest handing it out shot by shot instead. Only a small subset of syndromes gains much from the extra search, and the decoder can see which ones from the inside, before the extra work is done.
What $\dH$ picks up is not how bad a syndrome looks but a conflict inside the decoder: how far the constraints of the code override the explanation BP prefers. Two estimates built under different rules part company where that conflict is strongest, and those are the shots the fast path is most likely to get wrong. They are a minority at every operating point we measured, which is why a small escalation budget recovers most of what the full sweep would deliver.
The residual comparator makes this concrete, and it is the reason we included it. Three numbers come out of the same decoding pass and say progressively more: the syndrome weight counts how many detectors fired, the residual counts how many checks BP still fails, and the disagreement counts how many mechanisms have to change before it stops failing them. Going from the first to the second helps a great deal, and going from the second to the third helps a great deal more, even though the last two are strongly correlated. Knowing how many checks are still violated is not enough to match the ranking $\dH$ gives, so the reconstruction is telling us something about the way BP conflicts with the code, and not only about how much.

The four experiments sit in different places on that map. On $[[144,12,12]]$ the signal ranks risk well and the gap between the fast path and the full sweep is wide, and the gain is large. On $[[72,12,6]]$ the signal still ranks risk well and the gap is narrow, so little is available to win. The radial instance pairs a well-ranked signal with a wide gap again, on a code built a different way, so the ordering carries across to a different detector geometry. On the hardware, the signal survives device noise while the stronger search has almost nothing to repair. The headroom changes across the four; the ability of the disagreement to rank risk does not.

Spotting a risky shot and being able to repair it are different things. The decoder can be confident that a shot is in trouble and still have nothing better to offer it, because the stronger search cannot correct what went wrong. The $[[72,12,6]]$ code and the hardware experiment are both like this: in each, $\dH$ ranks risk well and the stronger search recovers very little. Together, these two conditions decide whether the method is worth using on a given code at a given noise strength. The signal tells us which shots to spend more computation on, and the strength of the escalated search sets how much logical error rate that spending can win back.

There is a practical consequence for classical resources. If only a fraction $\fesc$ of shots reaches the expensive branch, that branch has to keep up with $\fesc$ of the syndrome stream rather than all of it, so the expensive decoder can be sized for a fraction of the load rather than for all of it~\cite{terhal2015review,battistel2023realtime}. The timings reported here are single-process, single-threaded measurements of exactly that.

For a BP-OSD decoder, the practical conclusion is a narrow one. The benefit of deeper search is concentrated in a small subset of shots, and the decoder already knows, from two estimates it computes anyway, which ones those are. Across the two high-headroom constructions tested here, running the expensive search on those shots alone recovers most of the accuracy the full sweep would deliver. On the primary bivariate bicycle timing benchmark, it does so at a fraction of the serial decoding cost, and the signal that makes the allocation possible remains predictive on hardware.

\section{\label{sec:methods}Methods}

\subsection{Codes and syndrome extraction}

Bivariate bicycle codes are built from two polynomials $A$ and $B$ over the ring of the torus $\mathbb{Z}_\ell \times \mathbb{Z}_m$, with parity-check matrices $H_X = [A \,|\, B]$ and $H_Z = [B^\top |\, A^\top]$ and $n = 2\ell m$ data qubits~\cite{bravyi2024bbcodes}. The simulation benchmarks use the $[[144,12,12]]$ code with $(\ell,m) = (12,6)$, $A = x^3 + y + y^2$ and $B = y^3 + x + x^2$, over $24$ rounds of syndrome extraction, and the $[[72,12,6]]$ code with $(\ell,m) = (6,6)$ and the same polynomials, over $12$ rounds. The hardware experiment uses a small instance of the same construction with $(\ell,m) = (3,2)$ and $A = B = x + y$, for which $\mathrm{rank}\,H_X = \mathrm{rank}\,H_Z = 5$ and hence $k = 2$, and for which exhaustive enumeration gives $d_X = d_Z = 2$; it shares the bicycle layout with the larger codes but neither their polynomials nor their distance. A shot is a logical failure if any of the $k$ logical $Z$ observables is misidentified.

The cross-family test of Sec.~\ref{sec:radial} uses an instance of the radial code family of Ref.~\cite{scruby2026radial}, taken from the parity-check matrices published with that work, with construction parameters $(r,s) = (3,5)$, $n = 90$ and $k = 8$. We refer to this instance by $n$ and $k$; its distance was not independently established here. Checks have weight $6$ and qubits degree $3$, $H_X H_Z^\top = 0$ over $\mathbb{F}_2$, and the logical operators satisfy the CSS pairing conditions. The circuit is a twelve-round circuit-level $Z$-memory of the same form as the bivariate bicycle circuits, sampled at $p = 0.008$ and $p = 0.009$ with $20{,}000$ test shots per operating point and a further $5{,}000$ independent shots per point used only for threshold calibration.

\subsection{Noise model}

Circuits are simulated in Stim~\cite{gidney2021stim} with a uniform circuit-level depolarising model parameterised by a single scalar $p$: an \texttt{X\_ERROR}$(p)$ after every data and ancilla reset, a \texttt{DEPOLARIZE2}$(p)$ after every two-qubit gate, a \texttt{DEPOLARIZE1}$(p)$ on idling data qubits after each round, and a measurement flip probability $p$ on syndrome and final readout. Gates, idling and readout share the same $p$, with reset modelled as a bit flip rather than as full single-qubit depolarising. Decoding uses the detector error model obtained from the circuit with error decomposition enabled, which supplies $H$, the prior probabilities $p_i$ and the logical observables.

\subsection{Decoders}

BP and OSD-0 use the \texttt{ldpc} package, version 2.4.1~\cite{roffe2020decoding,roffe2022ldpc}, in the normalised min-sum formulation with a scaling factor of $0.625$, a parallel (flooding) update schedule, a maximum of $20$ iterations and a single OpenMP thread. The fast path returns the syndrome-consistent OSD-0 output. When the BP hard decision already satisfies the syndrome that output is the BP solution itself; otherwise the order-zero reconstruction sets the free bits to zero, selects an independent column set of size $\mathrm{rank}\,H$ greedily under the implementation's reliability ordering, and solves the resulting system over $\mathbb{F}_2$. For the $[[144,12,12]]$ detector error model $H$ has shape $1800 \times 12240$ with $\mathrm{rank}\,H = 1794$, leaving $k_{\mathrm{free}} = 10446$ free columns. Before it can score a single candidate, the escalated branch needs four things for that shot: the BP reliability ordering, an independent column set of size $\mathrm{rank}\,H$, the $\mathbb{F}_2$ solve that turns it into the order-zero solution, and the ordered list of remaining free columns that the sweep runs over. Our implementation derives all four itself and checks the resulting information set and order-zero solution bit for bit against the library output, rather than reusing the elimination carried out inside the library fast path, which the package does not expose. That re-derivation is what the algebraic preprocessing time reported in Sec.~\ref{sec:timing} pays for, and it is charged in full to every escalated shot. The truncated and full 1FV branches then share that ordering and the candidate score $S(x) = \sum_{i : x_i = 1} \log(1/p_i)$ evaluated on the detector error model channel probabilities, with the lowest-scoring candidate returned; $K = 1000$ and $K = 500$ search the first $1000$ and $500$ entries of the ordering, so the truncated branches, the full sweep and the adaptive policy are all timed on the same implementation. The radial detector error model has $k_{\mathrm{free}} = 3289$, so $K = 1000$ is a truncation there as well, and the same BP settings, fast-path convention and candidate score are used throughout. The score is a prior-weighted surrogate, not an exact independent-Bernoulli negative log likelihood.

\subsection{Routing and evaluation}

The router escalates the shots with the largest $\dH$. In the main figures the escalation budget is imposed exactly, by taking the top $k = \mathrm{round}(\fesc n)$ shots of the evaluation sample, so that accuracy and cost are reported on identical shots; the held-out threshold calibration is described in Appendix~\ref{app:controls}. The syndrome weight control ranks by $\lVert s \rVert_0$, the residual control by $r_{\mathrm{BP}} = \lVert H\eBP \oplus s \rVert_0$, and the random control draws the escalated subset uniformly, averaged over $200$ seeds.

The two radial operating points were fixed from a blind pilot that examined only the strong-search headroom, before any disagreement statistic, residual statistic or routing outcome was computed on that code. Calibrating a threshold on the $5{,}000$ separate radial shots and applying it to the $20{,}000$-shot test set gives realised escalation fractions of $22.13\%$ and $19.09\%$ against a nominal $20\%$, the difference coming from ties in the discrete values of $\dH$; the matched-budget comparisons in Sec.~\ref{sec:radial} use exact top-$k$ routing.

\subsection{Statistics}

Logical error rates, recovered fractions and AUROC values are reported with nonparametric bootstrap $95\%$ intervals, using $800$ resamples and the $2.5$ and $97.5$ percentiles.

\subsection{Timing protocol}

Timings were measured on a single compute node with Intel Xeon Platinum 8352Y processors at $2.20$~GHz, in one process with a single OpenMP thread and no GPU acceleration at any stage of decoding. The benchmark uses $1000$ shots at $p = 0.007$ with a fixed seed. The escalated branch terminates after its $K$ candidates rather than completing the sweep and prorating, so the $K = 500$ and $K = 1000$ costs are measured directly. The per-shot time includes the complete fast path together with the additional search on escalated shots. Absolute times are node dependent; the comparisons between the fast path, the adaptive policy and the fixed strong references are the quantities of interest.

\subsection{Hardware experiment}

The experiment was executed on the Quantinuum H2 trapped-ion quantum charge-coupled device processor~\cite{moses2023h2,pino2021qccd}. Hardware measurement outcomes were reconstructed from the named classical registers in the circuit measurement order and converted to detector events and logical observables using the corresponding Stim measurement-to-detector map, so that hardware and simulated records share identical detector and logical definitions. The Stim preflight uses the noise model supplied with the device description. Decoding and all analysis were performed offline in the HPC environment, using the same BP-OSD pipeline as the simulation study, with the endpoint fixed in advance.

\begin{acknowledgments}
The authors acknowledge funding support from the UK Engineering and Physical Sciences Research Council (EPSRC) under the project ``Software Environment for Actionable \& VVUQ-evaluated Exascale Applications (SEAVEA)'' (Grant No. EP/W007711/1). The authors thank Quantinuum for providing access to the H2 quantum processor, and the members of the Centre for Computational Science at University College London for their input during this research.
\end{acknowledgments}

\section*{Data and code availability}

The decoding and analysis code, together with the data supporting the findings of this study, are available at \url{https://github.com/UCL-CCS/Adaptive-qLDPC}.

\appendix

\section{\label{app:controls}Control experiments}

\subsection{BP convergence as a router}

Table~\ref{tab:bpflag} reports, for both codes, the BP convergence rate, how the OSD-0 failures split between converged and non-converged shots, and how well the convergence flag and $\dH$ each separate the failures from the successes. On the $[[144,12,12]]$ code no failure occurs among converged shots, and the flag says almost nothing about whether OSD-0 will fail. On the $[[72,12,6]]$ code it does carry information, and is still much weaker than $\dH$. The final column gives $\dH$ restricted to non-converged shots on the $[[144,12,12]]$ code.

\begin{table}[!htbp]
\caption{\label{tab:bpflag}BP convergence against BP-OSD disagreement, both scored on whether OSD-0 fails. Logical failures are split by whether BP converged on that shot. The last two columns are AUROC values for the binary convergence flag and for $\dH$.}
\begin{tabular*}{\columnwidth}{@{\extracolsep{\fill}}S[table-format=1.3] S[table-format=2.2] S[table-format=4.0] S[table-format=4.0] S[table-format=1.3] S[table-format=1.3]@{}}
\toprule
 & {BP conv.} & \multicolumn{2}{c}{OSD-0 failures} & \multicolumn{2}{c}{AUROC} \\
\cmidrule(lr){3-4}\cmidrule(lr){5-6}
{$p$} & {(\%)} & {conv.} & {non-conv.} & {BP flag} & {$\dH$} \\
\midrule
\multicolumn{6}{l}{$[[144,12,12]]$, $20{,}000$ shots per point} \\
0.006 & 9.18 & 0 & 95 & 0.546 & 0.958 \\
0.007 & 3.06 & 0 & 398 & 0.516 & 0.958 \\
0.008 & 0.68 & 0 & 1255 & 0.504 & 0.937 \\
\addlinespace
\multicolumn{6}{l}{$[[72,12,6]]$, $10{,}000$ shots per point} \\
0.006 & 54.33 & 46 & 360 & 0.724 & 0.831 \\
0.008 & 29.31 & 50 & 1289 & 0.648 & 0.839 \\
\bottomrule
\end{tabular*}
\end{table}

\subsection{Comparators for risk ranking}

Table~\ref{tab:comparators} collects the three decoder-internal comparators on the matched $[[144,12,12]]$ shots: the disagreement $\dH$, the BP residual $r_{\mathrm{BP}}$ of Eq.~\eqref{eq:res}, and the syndrome weight. The upper block gives how well each separates OSD-0 failures from successes, together with the Spearman correlation between $\dH$ and $r_{\mathrm{BP}}$; the lower block gives the logical error rate under exact top-$20\%$ routing to the $K = 1000$ branch.

\begin{table}[!htbp]
\caption{\label{tab:comparators}Risk-ranking comparators on the $[[144,12,12]]$ code, $20{,}000$ shots per operating point. The upper block gives the AUROC for OSD-0 failure and the Spearman correlation $\rho$ between $\dH$ and $r_{\mathrm{BP}}$. The lower block gives the logical error rate under exact top-$20\%$ routing.}
\begin{tabular*}{\columnwidth}{@{\extracolsep{\fill}}S[table-format=1.3] S[table-format=1.5] S[table-format=1.5] S[table-format=1.5] S[table-format=1.3]@{}}
\toprule
{$p$} & {$\dH$} & {$r_{\mathrm{BP}}$} & {$w_s$} & {$\rho$} \\
\midrule
\multicolumn{5}{l}{AUROC for OSD-0 failure} \\
0.006 & 0.958 & 0.842 & 0.654 & 0.956 \\
0.007 & 0.958 & 0.795 & 0.657 & 0.943 \\
0.008 & 0.937 & 0.769 & 0.659 & 0.927 \\
\addlinespace
\multicolumn{5}{l}{Logical error rate at $\fesc = 0.20$} \\
0.006 & 0.00135 & 0.00220 & 0.00365 & {} \\
0.007 & 0.00545 & 0.01140 & 0.01385 & {} \\
0.008 & 0.02530 & 0.04170 & 0.04840 & {} \\
\bottomrule
\end{tabular*}
\end{table}

\subsection{Hardware transition counts}

Table~\ref{tab:h2trans} gives the joint outcome of the fast path and the full 1FV sweep over the $300$ H2 shots. The sweep rescued three OSD-0 failures and introduced three, so the logical error rate is unchanged.

\begin{table}[!htbp]
\caption{\label{tab:h2trans}Joint outcomes of the fast path and the full 1FV sweep on the $300$ H2 hardware shots. The off-diagonal entries count three rescues and three shots turned into logical failures, so the logical error rate is unchanged.}
\centering
\setlength{\tabcolsep}{12pt}
\begin{tabular}{l S[table-format=3.0] S[table-format=3.0]}
\toprule
 & \multicolumn{2}{c}{full 1FV} \\
\cmidrule(lr){2-3}
OSD-0 & {correct} & {fail} \\
\midrule
correct & 214 & 3 \\
fail & 3 & 80 \\
\bottomrule
\end{tabular}
\end{table}

\subsection{Held-out threshold calibration}

The threshold $\tau$ was calibrated on twenty independent seeds as the $k$-th largest value of $\dH$ and applied as $\dH \geq \tau$ to twenty held-out seeds, with the folds then exchanged. Table~\ref{tab:holdout} reports the two-fold average at a nominal budget of $20\%$, alongside an exact top-$k$ reference evaluated separately within each $10{,}000$-shot fold. That reference is not numerically identical to the pooled top-$k$ value quoted in Sec.~\ref{sec:escalation}, which is computed over all $20{,}000$ shots at once.

\begin{table}[!htbp]
\caption{\label{tab:holdout}Held-out threshold calibration on the $[[144,12,12]]$ code at a nominal $20\%$ escalation budget. The fold-wise reference is the exact top-$k$ result computed separately within each $10{,}000$-shot calibration fold.}
\begin{tabular*}{\columnwidth}{@{\extracolsep{\fill}}S[table-format=1.3] S[table-format=2.0] S[table-format=2.1] S[table-format=1.5] S[table-format=1.5]@{}}
\toprule
 & & & \multicolumn{2}{c}{Logical error rate} \\
\cmidrule(lr){4-5}
{$p$} & {$\tau$} & {$\fesc$ (\%)} & {held-out} & {fold-wise top-$k$} \\
\midrule
0.006 & 14 & 20.9 & 0.00135 & 0.00135 \\
0.007 & 21 & 21.7 & 0.00525 & 0.00540 \\
0.008 & 32 & 20.8 & 0.02475 & 0.02505 \\
\bottomrule
\end{tabular*}
\end{table}

\section{\label{app:extended}Extended data}

\subsection{The $[[72,12,6]]$ code}

Table~\ref{tab:bb72} collects the logical error rates behind Fig.~\ref{fig:pareto}(d, e) for the low-headroom control, including the escalation points at $20\%$ and $30\%$ discussed in Sec.~\ref{sec:headroom}.

\begin{table}[!htbp]
\caption{\label{tab:bb72}Logical error rates for the $[[72,12,6]]$ code over $12$ rounds, $10{,}000$ shots per operating point. Escalation columns route the top $20\%$ and $30\%$ of shots by $\dH$ to the $K=1000$ branch. Headroom is the OSD-0 rate divided by the full 1FV rate.}
\begin{tabular*}{\columnwidth}{@{\extracolsep{\fill}}S[table-format=1.3] S[table-format=1.4] S[table-format=1.4] S[table-format=1.4] S[table-format=1.4] S[table-format=1.2]@{}}
\toprule
 & \multicolumn{4}{c}{Logical error rate} & \\
\cmidrule(lr){2-5}
{$p$} & {OSD-0} & {full 1FV} & {$\fesc = 0.2$} & {$\fesc = 0.3$} & {headroom} \\
\midrule
0.006 & 0.0406 & 0.0350 & 0.0358 & 0.0353 & 1.16 \\
0.008 & 0.1339 & 0.1101 & 0.1140 & 0.1121 & 1.22 \\
\bottomrule
\end{tabular*}
\end{table}

\subsection{Sensitivity to the truncated search depth}

Tables~\ref{tab:always} and~\ref{tab:ksens} compare the primary $K = 1000$ branch with a shallower $K = 500$ branch on the same matched $[[144,12,12]]$ shots, under the exact top-$k$ routing convention used in the main text. Applied to every shot, the shallower branch already recovers $82\%$ to $89\%$ of the improvement available between OSD-0 and the full 1FV reference, against $96\%$ to $97\%$ for $K = 1000$. With routing on disagreement the ordering holds at every budget, so the method does not depend on the particular choice $K = 1000$. The two branches are not equivalent: at $p = 0.007$ and $p = 0.008$ the deeper branch has appreciably more corrective capacity, and recovers a correspondingly larger share of the available improvement. The disagreement decides where search is allocated, and the strength of the escalated branch sets how much that allocation can deliver.

\begin{table}[!htbp]
\caption{\label{tab:always}Logical error rates when the search is applied to every shot, $[[144,12,12]]$ code, $20{,}000$ shots per operating point.}
\begin{tabular*}{\columnwidth}{@{\extracolsep{\fill}}S[table-format=1.3] S[table-format=1.5] S[table-format=1.5] S[table-format=1.5] S[table-format=1.5]@{}}
\toprule
 & \multicolumn{4}{c}{Logical error rate} \\
\cmidrule(lr){2-5}
{$p$} & {OSD-0} & {$K = 500$} & {$K = 1000$} & {full 1FV} \\
\midrule
0.006 & 0.00475 & 0.00145 & 0.00115 & 0.00105 \\
0.007 & 0.01990 & 0.00685 & 0.00475 & 0.00410 \\
0.008 & 0.06275 & 0.02645 & 0.02040 & 0.01865 \\
\bottomrule
\end{tabular*}
\end{table}

\begin{table}[!htbp]
\caption{\label{tab:ksens}Escalation on disagreement with $K = 500$ and $K = 1000$ on the $[[144,12,12]]$ code, under exact top-$k$ routing. LER is the logical error rate, and recovered is the fraction of the improvement available between OSD-0 and the full 1FV reference.}
\begin{tabular*}{\columnwidth}{@{\extracolsep{\fill}}S[table-format=1.3] S[table-format=1.1] S[table-format=1.5] S[table-format=2.1] S[table-format=1.5] S[table-format=2.1]@{}}
\toprule
 & & \multicolumn{2}{c}{$K = 500$} & \multicolumn{2}{c}{$K = 1000$} \\
\cmidrule(lr){3-4}\cmidrule(lr){5-6}
{$p$} & {$\fesc$} & {LER} & {rec.\ (\%)} & {LER} & {rec.\ (\%)} \\
\midrule
0.006 & 0.1 & 0.00185 & 78.4 & 0.00160 & 85.1 \\
0.006 & 0.2 & 0.00160 & 85.1 & 0.00135 & 91.9 \\
0.006 & 0.3 & 0.00145 & 89.2 & 0.00120 & 95.9 \\
\addlinespace
0.007 & 0.1 & 0.00915 & 68.0 & 0.00710 & 81.0 \\
0.007 & 0.2 & 0.00755 & 78.2 & 0.00545 & 91.5 \\
0.007 & 0.3 & 0.00710 & 81.0 & 0.00500 & 94.3 \\
\addlinespace
0.008 & 0.1 & 0.03800 & 56.1 & 0.03295 & 67.6 \\
0.008 & 0.2 & 0.03095 & 72.1 & 0.02530 & 84.9 \\
0.008 & 0.3 & 0.02810 & 78.6 & 0.02215 & 92.1 \\
\bottomrule
\end{tabular*}
\end{table}

\FloatBarrier
\subsection{Radial-code cross-family validation}

Tables~\ref{tab:radial} and~\ref{tab:radialsig} give the formal results for the $n = 90$, $k = 8$ radial instance of Sec.~\ref{sec:radial}, from independent test sets of $20{,}000$ shots at each operating point.

\begin{table}[!htbp]
\caption{\label{tab:radial}Fixed policies on the $n = 90$, $k = 8$ radial instance, $20{,}000$ test shots per operating point.}
\begin{tabular*}{\columnwidth}{@{\extracolsep{\fill}}S[table-format=1.3] S[table-format=1.5] S[table-format=1.5] S[table-format=1.5]@{}}
\toprule
 & \multicolumn{3}{c}{Logical error rate} \\
\cmidrule(lr){2-4}
{$p$} & {OSD-0} & {$K = 1000$} & {full 1FV} \\
\midrule
0.008 & 0.05130 & 0.02095 & 0.02090 \\
0.009 & 0.10160 & 0.04660 & 0.04655 \\
\bottomrule
\end{tabular*}
\end{table}

\begin{table*}[!t]
\caption{\label{tab:radialsig}Risk ranking and exact top-$20\%$ routing on the $n = 90$, $k = 8$ radial instance. Every method escalates the same number of shots to the same $K = 1000$ branch, so the comparison is at matched budget. Recovered is the fraction of the OSD-0 to full-1FV improvement obtained by routing on $\dH$; the bootstrap $95\%$ intervals on $\dH$ are quoted in Sec.~\ref{sec:radial}.}
\centering
\setlength{\tabcolsep}{14pt}
\begin{tabular}{S[table-format=1.3] S[table-format=1.5] S[table-format=1.5] S[table-format=1.5] S[table-format=1.5] S[table-format=2.1]}
\toprule
{$p$} & {$\dH$} & {$r_{\mathrm{BP}}$} & {$w_s$} & {random} & {rec.\ (\%)} \\
\midrule
\multicolumn{6}{l}{AUROC for OSD-0 failure} \\
0.008 & 0.951 & 0.832 & 0.707 & {} & {} \\
0.009 & 0.940 & 0.809 & 0.701 & {} & {} \\
\addlinespace
\multicolumn{6}{l}{Logical error rate at $\fesc = 0.20$} \\
0.008 & 0.02275 & 0.03290 & 0.03860 & 0.04525 & 93.9 \\
0.009 & 0.05390 & 0.07275 & 0.07975 & 0.09054 & 86.6 \\
\bottomrule
\end{tabular}
\end{table*}

\bibliographystyle{apsrev4-2}
\bibliography{adaptive}

\end{document}